\documentclass[12pt]{iopart}

\usepackage{iopams}  
\usepackage{cite}
\usepackage{cleveref}
\usepackage{hyperref}
\usepackage{lipsum}

\renewcommand{\theequation}{\thesection.\arabic{equation}}
\renewcommand{\theenumi}{(\thesection.\alph{enumi})}

\begin{document}

\title[Structural equations of supermanifolds immersed in $\mathcal{M}^{(3\vert2)}(c)$]{Structural equations of supermanifolds immersed in the superspace $\mathcal{M}^{(3\vert2)}(c)$ with a prescribed curvature}

\author{S Bertrand}
\address{Czech Technical University in Prague, Faculty of Nuclear Sciences and Physical Engineering, Department of Physics, B\v{r}ehov\'a 7, 115 19 Prague 1, Czech Republic}
\ead{bertrseb@fjfi.cvut.cz}

\author{A M Grundland}
\address{Centre de Recherches Math\'ematiques, Universit\'e de Montr\'eal,\\ CP 6128 Succ. Centre-Ville, Montr\'eal (Qc) H3C 3J7, Canada}
\vspace{3mm}
\address{Department of Mathematics and Computer Science,\\ Universit\'e du Qu\'ebec \`a Trois-Rivi\`eres,\\ 3351, boul. des Forges CP 500, Trois-Rivi\`eres (Qc) G9A 5H7, Canada}
\ead{grundlan@crm.umontreal.ca}

\begin{abstract}
The aim of this paper is to construct the structural equations of supermanifolds immersed in Euclidean, hyperbolic and spherical superspaces parametrised with two bosonic and two fermionic variables. To perform this analysis, for each type of immersion, we split the supermanifold into its Grassmannian components and study separately each manifold generated. Even though we consider four variables in the Euclidean case, we obtain that the structural equations of each manifold are linked with the Gauss--Codazzi equations of a surface immersed in a Euclidean or spherical space. In the hyperbolic and spherical superspaces, we find that the body manifolds are linked with the classical Gauss--Codazzi equations for a surface immersed in hyperbolic and spherical spaces, respectively. For some soul manifolds, we show that the immersion of the manifolds must be in a hyperbolic space and that the structural equations split into two cases. In one case, the structural equations reduce to the Liouville equation, which can be completely solved. In the other case, we can express the geometric quantities solely in terms of the metric coefficients, which provide a geometric characterization of the structural equations in terms of functions linked with the Hopf differential, the mean curvature and a new function which does not appear in the characterization of a classical (not super) surface.
\end{abstract}


\vspace{2pc}


\maketitle

\section{Introduction}\label{SecIntro}\setcounter{equation}{0}
The fundamental ideas of the Enneper and Weierstrass formula for an immersion of minimal surfaces in three-dimensional Euclidean space \cite{Enneper,Weierstrass} have been intensively developed with the purpose of extending this construction for obtaining more general types of immersions of surfaces. A review of these efforts can be found in many books, e.g. \cite{BE00,Cartan53,Helein01,RS02,NS94}. An interesting link between the theory of surfaces defined by a moving frame and that of integrable systems (soliton theory) was pursued, leading to numerous applications \cite{Uhlenbeck89,Sym83,Konopelchenko96,Guest97,GGP12,GP11,GG10,FGFL,FG96}. Surfaces immersed in Lie groups, Lie algebras and homogeneous spaces appear in many areas of physics, chemistry and biology (see e.g.  \cite{Davydov79,MS04,Mikhailov86,Rajaraman02,Safran94,Landolfi03,JMN14,GPW91,PS91} and references therein). The algebraic approach to the structural equations of these surfaces proved to be fruitful from the point of view of constructing surfaces in closed form. Therefore, it seems to be worthwhile to try to extend this method and check its effectiveness for the case of supermanifolds immersed in Euclidean, spherical and hyperbolic superspaces. This is in short the aim of the paper.

Techniques for obtaining a representation of integrable equations in the zero-curvature form associated with the Gauss--Weingarten equations (which are linear differential equations) are much better understood in the classical case than for their supersymmetric extensions. This is particularly true, for instance, in the case of the inverse scattering transformation and for classical integration techniques using moving frames on surfaces which lie in the Lie algebras/groups \cite{FGFL}. For partial differential equations (PDEs) with two independent variables written in the zero-curvature form
\begin{equation}
U_y-V_x+[U,V]=0,\label{0ZCC}
\end{equation}
the matrices $U$ and $V$ are considered to be integrable in the sense that these matrices can be extended in a non-trivial way to a one-parameter family $U(x,y,\lambda)$, $V(x,y,\lambda)$ satisfying (\ref{0ZCC}), such that the integrable set of PDEs is preserved. In soliton theory, these matrices are usually rational functions of the spectral parameter $\lambda$. However for equations coming from differential geometry, this parameter $\lambda$ does not appear in the Gauss--Weingarten equations,
\begin{equation}
\Phi_x=U(x,y)\Phi,\qquad \Phi_y=V(x,y)\Phi.
\end{equation}
This fact makes integrable equations an interesting subject of study and are related to special classes of surfaces \cite{BE00}. In this context, the problem of constructing and investigating (by analytical methods) the structural equations of supermanifolds immersed in superspaces with constant prescribed curvature is closely related to the study of supergroup properties of differential equations. The fact that integrable equations occur in surface theory with $\lambda$ deformations makes it possible to construct a regular algorithm for finding certain classes of surfaces within $\lambda$--described deformations without invoking any additional considerations but proceeding only from the frame on the surface and their compatibility conditions (i.e. the Gauss--Codazzi equations).

The objective of this paper is to investigate the structural equations of Riemannian supermanifolds immersed in a Euclidean superspace, in a hyperbolic superspace and in a spherical superspace, all denoted by $\mathcal{M}^{(3\vert2)}(c)$, with two bosonic independent variables and two fermionic independent variables. To perform this analysis, we define that the superspace is equipped with an inner product of the form
\begin{equation}
\langle A\vert B\rangle=\langle A_0,B_0\rangle+\langle A_3,B_3\rangle\xi_1\xi_2+\langle A_2,B_2\rangle\xi_1\xi_3+\langle A_1,B_1\rangle\xi_2\xi_3,
\end{equation} 
where the $\xi_k$, $k=1,2,3$, are Grassmann numbers,
\begin{eqnarray*}
A=A_0+A_3\xi_1\xi_2+A_2\xi_1\xi_3+A_1\xi_2\xi_3,\\
B=B_0+B_3\xi_1\xi_2+B_2\xi_1\xi_3+B_1\xi_2\xi_3,\\
A_i,B_i\in\mathbb{R}^{N_i}(\mbox{or }\mathbb{C}^{N_i}),\qquad i=0,1,2,3\nonumber
\end{eqnarray*}
and the bilinear product $\langle\cdot,\cdot\rangle$ is either the usual Euclidean inner product
\begin{equation}
\langle A_i,B_i\rangle=\sum_{j=1}^{N_i}A_{ij}B_{ij}
\end{equation}
for the Euclidean superspace or the usual hyperbolic/spherical inner product
\begin{eqnarray}
\langle A_i,B_i\rangle=\mbox{sgn}(c_i)A_{i0}B_{i0}+\sum_{j=1}^{N_i}A_{ij}B_{ij},\\
\langle A_i,A_i\rangle=\langle B_i,B_i\rangle=c_i\in\mathbb{R}\nonumber
\end{eqnarray}
for the hyperbolic/spherical superspace. This form of inner product allows us to split the supermanifolds into manifolds for each coefficient of the Grassmann numbers. For each type of manifold, we investigate its structural equations through a moving frame formalism, which are unique up to Euclidean motions in the Euclidean/hyperbolic/spherical space. The resulting set of PDEs are to be solved or are to be reduced to a lower number of PDEs using solely derivatives with respect to the bosonic independent variables.

The paper is organized as follow. In section \ref{SecEuc}, we consider a supermanifold immersed in a Euclidean superspace, which can be decomposed into three types of manifold. In subsection \ref{SecF0}, we consider a manifold immersed in a Euclidean space involving only the bosonic independent variables. In subsection \ref{SecF2}, we consider the manifolds immersed in a Euclidean space involving two bosonic and one fermionic independent variables. In subsection \ref{SecF3}, we consider the manifolds immersed in a Euclidean space involving all four independent variables. In section \ref{SecSH}, we consider supermanifolds immersed in a hyperbolic superspace or in a spherical superspace, which are decomposed into three types of manifold immersed in a hyperbolic space or in a spherical space, respectively. In subsection \ref{SecNEF0}, we consider the manifolds immersed in a hyperbolic/spherical space involving only the bosonic independent variables. In subsection \ref{SecNEF2}, we consider the manifolds immersed in a hyperbolic/spherical space involving two bosonic and one fermionic independent variables. In subsection \ref{SecNEF3}, we consider the manifolds immersed in a hyperbolic/spherical space involving all four independent variables. In section \ref{SecConc}, we provide conclusions and some future perspectives.

\section{Structural equations of a supermanifold in a Euclidean superspace}\label{SecEuc}\setcounter{equation}{0}
Let $\mathcal{S}$ be a bosonic-valued Riemannian supermanifold parametrised through two bosonic independent variables $x_+$ and $x_-$ together with two fermionic independent variables $\theta^+$ and $\theta^-$. In this paper, we consider three$^{\hyperlink{foo}{1}}$ ~(fermionic) Grassmann numbers $\lbrace\xi_1,\xi_2,\xi_3\rbrace$.  We assume that the variables $x_+$ and $x_-$ possess a non-zero body component and that $\theta^+$ can be expressed as the product of the Grassmann generator $\xi_1$ with a bosonic variable $\theta_+$ and analogously for $\theta^-$ with $\xi_2$, i.e.
\begin{equation}
\theta^+=\xi_1\theta_+,\qquad\theta^-=\xi_2\theta_-,
\end{equation}
where $\theta_+$ and $\theta_-$ possess a non-zero body part.

In order to investigate the structural equations of the supermanifold $\mathcal{S}$ and its geometric properties, we split the supermanifold $\mathcal{S}$ into a power series of Grassmann numbers for which each coefficient represents a manifold with new real (or complex) variables, e.g.
\begin{equation}
\hspace{-2cm}\mathcal{S}_F=F_0(x_1,x_2)+\xi_1\xi_2F_3(x_1,x_2,\theta_3,\theta_4)+\xi_1\xi_3F_2(x_1,x_2,\theta_3)+\xi_2\xi_3F_1(x_1,x_2,\theta_4),\label{Fdecom}
\end{equation}
where the new variables $x_1$, $x_2$, $\theta_3$ and $\theta_4$ are associated with $x_+$, $x_-$, $\theta_+$ and $\theta_-$, respectively. Throughout this paper, we use the abbreviated notation for the partial derivatives,
\begin{equation}
\partial_1=\frac{\partial}{\partial x_1},\qquad\partial_2=\frac{\partial}{\partial x_2},\qquad\partial_3=\frac{\partial}{\partial \theta_3},\qquad\partial_4=\frac{\partial}{\partial \theta_4}.
\end{equation}
Each manifold $F_i$, $i=0,1,2,3$, must satisfy the relations
\begin{equation}
\partial_3^2F_i=\partial_4^2F_i=0
\end{equation}
due to the properties of the fermionic variables $\theta^+$ and $\theta^-$. We assume throughout this paper that all manifolds are orientable and sufficiently smooth. 

We define the inner product $\langle\cdot\vert\cdot\rangle$ on the supermanifold $\mathcal{S}_F$ to be the sum of the inner products $\langle\cdot,\cdot\rangle$ on each manifold $F_i$, $i=0,1,2,3$, times their corresponding polynomial combinations of Grassmann numbers, i.e.
\begin{equation}
\langle \mathcal{S}_F\vert \mathcal{S}_F\rangle=\langle F_0,F_0\rangle+\langle F_3,F_3\rangle\xi_1\xi_2+\langle F_2,F_2\rangle\xi_1\xi_3+\langle F_1,F_1\rangle\xi_2\xi_3,
\end{equation}
where the inner product $\langle\cdot,\cdot\rangle$ on each manifold $F_i$, $i=0,1,2,3$, is the classical scalar product, e.g. for two vectors $A,B\in\mathbb{C}^3$
\begin{eqnarray}
\langle A,B\rangle=A_1B_1+A_2B_2+A_3B_3,\\ 
A=(A_1,A_2,A_3)^T,\qquad B=(B_1,B_2,B_3)^T,\nonumber
\end{eqnarray}
\rule{3cm}{.1mm}\\
\begin{footnotesize}
\hypertarget{foo}{1. } It would be straightforward to consider additional Grassmann numbers (as described later).
\end{footnotesize}
\pagebreak

\noindent where $(\cdot)^T$ stands for the transpose of the vector. Hence, the metric is represented by the $4\times4$ symmetric matrix
\begin{eqnarray}
\hspace{-1cm}g&=&[\langle\partial_j\mathcal{S}_F\vert\partial_k\mathcal{S}_F\rangle]\qquad\qquad j,k=1,2,3,4,\\
\hspace{-1cm}&=&[\langle\partial_jF_0,\partial_kF_0\rangle]+[\langle\partial_jF_3,\partial_kF_3\rangle]\xi_1\xi_2+[\langle\partial_jF_2,\partial_kF_2\rangle]\xi_1\xi_3+[\langle\partial_jF_1,\partial_kF_1\rangle]\xi_2\xi_3,\nonumber
\end{eqnarray}
where $\partial_j\mathcal{S}_F$ represent the tangent vectors to the supermanifold $\mathcal{S}_F$ and $\partial_jF_i$, $i=0,1,2,3$, represent the tangent vectors to the manifolds $F_i$. We assume that there exists a normal vector $N$ of the form
\begin{equation}
N=N_0+N_3\xi_1\xi_2+N_2\xi_1\xi_3+N_1\xi_2\xi_3,
\end{equation}
which satisfies the properties
\begin{equation}
\langle N\vert N\rangle=1+\xi_1\xi_2+\xi_1\xi_3+\xi_2\xi_3,\qquad\langle\partial_j\mathcal{S}_F\vert N\rangle=0,\quad j=1,2,3,4,
\end{equation}
i.e. such that all $N_i$, $i=0,1,2,3$, are unitary in their respective spaces.

As a result, we find that the supermanifold $\mathcal{S}_F$ consists of copies of three types of manifold ($F_1$ being a copy of $F_2$). At this point, it is clear that adding more Grassmann numbers $\xi_i$, $i>3$, would only add more copies of manifold with similar structural equations to those of $F_0$, $F_2$ or $F_3$. Therefore, we consider only the three types of manifolds $F_0$, $F_2$ and $F_3$. Moreover, this method can be applied for a fermionic supermanifold, but we still get similar structural equations as for $F_0$, $F_2$ and $F_3$. Hence, there is no need to explicitly present the fermionic version. One should note that for all three types of manifold, we consider the same conformal parametrisation ($x_1=\overline{x_2}$, $\theta_3=\overline{\theta_4}$) and we drop all unnecessary quantities related to the dependency in $\theta_3$ and $\theta_4$ or to the derivatives with respect to $\theta_3$ and $\theta_4$. In addition, to simplify the notation, in the three following subsections we will not keep the indices of the quantities which specify to which manifold ($F_0$, $F_2$ or $F_3$) they belong.

\subsection{Structural equations for a manifold of type $F_0$ in a Euclidean space}\label{SecF0}
For the surface associated with $F_0$, renamed $F$ in this subsection, the metric takes the form
\begin{equation}
\hspace{-2cm}g=[g_{ij}]=\left(\begin{array}{cc}
0 & \frac{1}{2}e^{u} \\
\frac{1}{2}e^{u} & 0
\end{array}\right),\qquad g_{ij}=\langle \partial_iF,\partial_jF\rangle,\qquad i,j=1,2,\label{g0}
\end{equation}
where $u$ is a real function of $x_1$ and $x_2$. We obtain the well-known geometry of a conformally parametrized surface immersed in a 3-dimensional Euclidean space. The moving frame on the surface is given by
\begin{equation}
\Omega=\left(\begin{array}{ccc}\partial_1F,&\partial_2F,&N\end{array}\right)^T\label{rep0}.
\end{equation}
Assuming that we can write the second derivatives of the immersion function $F$ and the first derivatives of the normal unit vector $N$ as a linear combination of the elements of the moving frame $\Omega$, i.e.
\begin{eqnarray}
\partial_i\partial_jF&=\sum_{k=1}^2\Gamma_{ij}^k\partial_kF+b_{ij}N,&\qquad i,j=1,2,\label{DDF0}\\
\hspace{4mm}\partial_iN&=\sum_{j=1}^{2}-b_{i}^j\partial_jF+\omega_{i}N,&\qquad i=1,2,\label{DN0}
\end{eqnarray}
we can construct the Gauss--Weingarten equations in the form
\begin{equation}
\partial_i\Omega=U_{i}\Omega,\qquad i=1,2,\label{GW0}
\end{equation}
where
\begin{equation}
U_{i}=\left(\begin{array}{ccc}\label{Ui0}
\Gamma_{i1}^1 & \Gamma_{i1}^2 & b_{i1} \\
\Gamma_{i2}^1 & \Gamma_{i2}^2 & b_{i2} \\
-b_{i}^1 & -b_{i}^2 & \omega_{i}
\end{array}\right).
\end{equation}
The coefficients $\Gamma_{ij}^k$ are the Christoffel symbols of second kind and the coefficients $b_{ij}$ are the coefficients of the second fundamental form. 
In addition, the Gauss--Codazzi equations are obtained through the compatibility condition of the Gauss--Weingarten equations (\ref{GW0}), i.e.
\begin{equation}
\partial_1U_{2}-\partial_2U_{1}+[U_{2},U_{1}]=0.\label{ZCC0}
\end{equation}
Explicitly, the Gauss--Weingarten equations are given by the matrices
\begin{eqnarray}
\hspace{-2cm}U_{1}=\left(\begin{array}{ccc}
\partial_1u & 0 & Q \\
0 & 0 & \frac{1}{2}He^{u} \\
-H & -2e^{-u}Q & 0
\end{array}\right),\label{U10}\qquad U_{2}=\left(\begin{array}{ccc}
0 & 0 & \frac{1}{2}He^{u} \\
0 & \partial_2u & \bar{Q} \\
-2e^{-u}\bar{Q} & -H & 0
\end{array}\right),\label{U20}
\end{eqnarray}
and the Gauss--Codazzi equations are
\begin{eqnarray}
\partial_1\partial_2u+\frac{1}{2}e^{u}H^2-2e^{-u}\vert Q\vert^2=0,\label{GC10}\\
\partial_2Q-\frac{1}{2}e^{u}\partial_1H=0,\qquad \partial_1\bar{Q}-\frac{1}{2}e^{u}\partial_2H=0,\label{GC20}
\end{eqnarray}
where $Q=Q(x_1,x_2)=\langle \partial_1^2F,N\rangle\in\mathbb{C}$ is associated with the Hopf differential and $H=H(x_1,x_2)=2e^{-u}\langle \partial_1\partial_2F,N\rangle\in\mathbb{R}$ is the mean curvature of the surface.

\subsection{Structural equations for a manifold of type $F_2$ in a Euclidean space}\label{SecF2}
For the manifold associated with $F_2$, renamed $F$ in this subsection, the metric takes the form
\begin{equation}
\hspace{-2cm}g=[g_{ij}]=\left(\begin{array}{ccc}
0 & \frac{1}{2}e^{\phi} & 0 \\
\frac{1}{2}e^{\phi} & 0 & 0 \\
0 & 0 & e^\psi
\end{array}\right),\qquad g_{ij}=\langle \partial_iF,\partial_jF\rangle,\qquad i,j=1,2,3,\label{g2}
\end{equation}
where $\psi$ is a real function of $x_1$ and $x_2$ and $\phi$ is a real function of $x_1$, $x_2$ and $\theta_3$ satisfying $\partial_3^3e^\phi=0$. One should note that $\theta_3$ has a zero imaginary part. The moving frame on the manifold is given by
\begin{equation}
\Omega=\left(\begin{array}{cccc}\partial_1F,&\partial_2F,& \partial_3F,&N\end{array}\right)^T.\label{rep2}
\end{equation}
The second derivatives of the immersion function $F$ and the first derivatives of the normal unit vector $N$ can be written as a linear combination of the elements of the moving frame $\Omega$, i.e.
\begin{eqnarray}
\partial_i\partial_jF&=\sum_{k=1}^3\Gamma_{ij}^k\partial_kF+b_{ij}N,&\qquad i,j=1,2,3,\label{DDF2}\\
\hspace{4mm}\partial_iN&=\sum_{j=1}^{3}-b_{i}^j\partial_jF+\omega_{i}N,&\qquad i=1,2,3,\label{DN2}
\end{eqnarray}
from where we can construct the Gauss--Weingarten equations in the form
\begin{equation}
\partial_i\Omega=U_{i}\Omega,\qquad i=1,2,3,\label{GW2}
\end{equation}
where
\begin{equation}
U_{i}=\left(\begin{array}{cccc}\label{Ui2}
\Gamma_{i1}^1 & \Gamma_{i1}^2 & \Gamma_{i1}^3 & b_{i1} \\
\Gamma_{i2}^1 & \Gamma_{i2}^2 & \Gamma_{i2}^3 & b_{i2} \\
\Gamma_{i3}^1 & \Gamma_{i3}^2 & \Gamma_{i3}^3 & b_{i3} \\
-b_{i}^1 & -b_{i}^2 & -b_{i}^3 & \omega_{i}
\end{array}\right).
\end{equation}
In addition, the Gauss--Codazzi equations are obtained through the compatibility condition of the Gauss--Weingarten equations (\ref{GW2}), i.e.
\begin{equation}
\partial_iU_{j}-\partial_jU_{i}+[U_{j},U_{i}]=0,\qquad i,j=1,2,3.\label{ZCC2}
\end{equation}

As a result, we get the set of PDEs
\begin{eqnarray}
\partial_1\partial_2\phi+\frac{1}{2}e^\phi\left(H^2+\frac{(\partial_3\phi)^2}{4k^2}\right)-2e^{-\phi}\vert Q\vert^2=0,\\
\partial_1\partial_3\phi=\partial_2\partial_3\phi=0,\qquad\partial_3^3e^\phi=0,\\
\partial_2Q-\frac{1}{2}e^\phi\partial_1H=0,\qquad\partial_1\bar{Q}-\frac{1}{2}e^\phi\partial_2H=0,\\
\partial_3Q-\frac{1}{2}Q\partial_3\phi=0,\qquad \partial_3\bar{Q}-\frac{1}{2}\bar{Q}\partial_3\phi=0,\\
2\partial_3^2\phi+(\partial_3\phi)^2=0,\qquad\partial_3H+\frac{1}{2}H\partial_3\phi=0.
\end{eqnarray}
By solving these equations, we find that the metric takes the form
\begin{equation}
\hspace{-1cm}g=\left(\begin{array}{ccc}
0 & \frac{1}{2}e^u(a\theta_3+b)^2 & 0 \\
\frac{1}{2}e^u(a\theta_3+b)^2 & 0 & 0 \\
0 & 0 & k^2
\end{array}\right),\qquad\begin{array}{l}
u=u(x_1,x_2)\in\mathbb{R},\\
a,b,k\in\mathbb{R}.
\end{array}\label{gsol2}
\end{equation}
The Gauss--Weingarten equations (\ref{GW2}) are given in terms of the matrices
\begin{eqnarray}
\hspace{-2.5cm}U_{1}=\left(\begin{array}{cccc}
\partial_1u & 0 & 0 & q(a\theta_3+b) \\
0 & 0 & -ae^u(a\theta_3+b)/2k^2 & \frac{1}{2}e^uh(a\theta_3+b) \\
a/(a\theta_3+b) & 0 & 0 & 0 \\
-h/(a\theta_3+b) & -2e^{-u}q/(a\theta_3+b) & 0 & 0
\end{array}\right),\label{U12}\\
\hspace{-2.5cm}U_{2}=\left(\begin{array}{cccc}
0 & 0 & -ae^u(a\theta_3+b)/2k^2 & \frac{1}{2}e^uh(a\theta_3+b) \\
0 & \partial_2u & 0 & \bar{q}(a\theta_3+b) \\
0 & a/(a\theta_3+b) & 0 & 0 \\
-2e^{-u}\bar{q}/(a\theta_3+b) & -h/(a\theta_3+b) & 0 & 0
\end{array}\right),\label{U22}\\
\hspace{-2.5cm}U_{3}=\left(\begin{array}{cccc}
a/(a\theta_3+b) & 0 & 0 & 0 \\
0 & a/(a\theta_3+b) & 0 & 0 \\
0 & 0 & 0 & 0 \\
0 & 0 & 0 & 0
\end{array}\right),\label{U32}
\end{eqnarray}
and the remaining compatibility conditions are the following Gauss--Codazzi equations:
\begin{eqnarray}
\partial_1\partial_2u+\frac{1}{2}e^{u}\left(h^2+\frac{a^2}{k^2}\right)-2e^{-u}\vert q\vert^2=0,\label{GC12}\\
\partial_2q-\frac{1}{2}e^{u}\partial_1h=0,\qquad \partial_1\bar{q}-\frac{1}{2}e^{u}\partial_2h=0,\label{GC22}
\end{eqnarray}
where 
\begin{eqnarray}
\langle\partial_1^2F,N\rangle=Q=q(x_1,x_2)(a\theta_3+b)\in\mathbb{C},\\
\langle\partial_2^2F,N\rangle=\bar{Q}=\bar{q}(x_1,x_2)(a\theta_3+b)\in\mathbb{C},\\
\langle\partial_1\partial_2F,N\rangle=\frac{1}{2}e^{\phi}H=\frac{1}{2}e^uh(x_1,x_2)(a\theta_3+b),\\
H=h(x_1,x_2)/(a\theta_3+b)\in\mathbb{R}.
\end{eqnarray}
Hence, the structural equations of the manifold $F_2$ are linked with the immersion of a surface in a spherical space $S^3(k^2/a^2)$.

\subsection{Structural equations for a manifold of type $F_3$ in a Euclidean space}\label{SecF3}
For the manifold associated with $F_3$, renamed $F$ in this subsection, the metric takes the form
\begin{equation}
\hspace{-2cm}g=[g_{ij}]=\left(\begin{array}{cccc}
0 & \frac{1}{2}e^{\phi} & 0 & 0 \\
\frac{1}{2}e^{\phi} & 0 & 0 & 0 \\
0 & 0 & 0 & \frac{1}{2}e^{\psi} \\
0 & 0 & \frac{1}{2}e^{\psi} & 0
\end{array}\right),\qquad \begin{array}{c}
g_{ij}=\langle \partial_iF,\partial_jF\rangle,\\
i,j=1,2,3,4,\end{array}\label{g3}
\end{equation}
where $\psi$ and $\phi$ are real functions of $x_1$, $x_2$ $\theta_3$ and $\theta_4$ satisfying $\partial_3^3e^\phi=\partial_4^3e^\phi=0$ and $\partial_3^2e^\psi=\partial_4^2e^\psi=0$. The moving frame on the manifold is given by
\begin{equation}
\Omega=\left(\begin{array}{ccccc}
\partial_1F, & \partial_2F, & \partial_3F, & \partial_4F, & N
\end{array}\right)^T.\label{rep3}
\end{equation}
Assuming that we can write the second derivatives of the immersion function $F$ and the first derivatives of the normal unit vector $N$ as a linear combination of the elements of the moving frame $\Omega$, i.e.
\begin{eqnarray}
\partial_i\partial_jF&=\sum_{k=1}^4\Gamma_{ij}^k\partial_kF+b_{ij}N,&\qquad i,j=1,2,3,4,\label{DDF3}\\
\hspace{4mm}\partial_iN&=\sum_{j=1}^{4}-b_{i}^j\partial_jF+\omega_{i}N,&\qquad i=1,2,3,4,\label{DN3}
\end{eqnarray}
we can construct the Gauss--Weingarten equations in the form
\begin{equation}
\partial_i\Omega=U_{i}\Omega,\qquad i=1,2,3,4,\label{GW3}
\end{equation}
where
\begin{equation}
U_{i}=\left(\begin{array}{ccccc}\label{Ui3}
\Gamma_{i1}^1 & \Gamma_{i1}^2 & \Gamma_{i1}^3 & \Gamma_{i1}^4 & b_{i1} \\
\Gamma_{i2}^1 & \Gamma_{i2}^2 & \Gamma_{i2}^3 & \Gamma_{i2}^4 & b_{i2} \\
\Gamma_{i3}^1 & \Gamma_{i3}^2 & \Gamma_{i3}^3 & \Gamma_{i3}^4 & b_{i3} \\
\Gamma_{i4}^1 & \Gamma_{i4}^2 & \Gamma_{i4}^3 & \Gamma_{i4}^4 & b_{i4} \\
-b_{i}^1 & -b_{i}^2 & -b_{i}^3 & -b_{i}^4 & \omega_{i}
\end{array}\right).
\end{equation}
In addition, the Gauss--Codazzi equations are obtained through the compatibility condition of the Gauss--Weingarten equations (\ref{GW3}), i.e.
\begin{equation}
\partial_iU_{j}-\partial_jU_{i}+[U_{j},U_{i}]=0,\qquad i,j=1,2,3,4.\label{ZCC3}
\end{equation}

As a result, we obtain a set of PDEs which splits into two cases:
\begin{enumerate}
\item $\partial_3\psi=\partial_4\psi=0$,\label{case1}
\item $\partial_3\phi=\partial_4\phi=0$.\label{case2}
\end{enumerate}
The explicit set of PDEs can be found in Appendix A.

\subsubsection{The case \ref{case1}, when $\partial_3\psi=\partial_4\psi=0$.}~\\
As a result for the case \ref{case1}, the metric takes the form
\begin{equation}
\hspace{-1cm}g=\left(\begin{array}{cccc}
0 & \frac{1}{2}e^u\omega^2 & 0 & 0 \\
\frac{1}{2}e^u\omega^2 & 0 & 0 & 0 \\
0 & 0 & 0 & \frac{1}{2}k^2\\
0 & 0 & \frac{1}{2}k^2 & 0
\end{array}\right),\qquad\begin{array}{l}
u=u(x_1,x_2)\in\mathbb{R},\\
\omega=\alpha\theta_3+\bar{\alpha}\theta_4+\gamma,\\
\gamma,k\in\mathbb{R},\qquad\alpha\in\mathbb{C}.
\end{array}\label{gsol31}
\end{equation}
The Gauss--Weingarten equations (\ref{GW3}) are given in terms of the matrices
\begin{eqnarray}
\hspace{-1cm}U_{1}=\left(\begin{array}{ccccc}
\partial_1u & 0 & 0 & 0 & q\omega \\
0 & 0 & -\bar{\alpha}\omega/k^2 & -\alpha\omega/k^2 & e^uh\omega/2 \\
\alpha/\omega & 0 & 0 & 0 & 0 \\
\bar{\alpha}/\omega & 0 & 0 & 0 & 0 \\
-h/\omega & -2e^{-u}q/\omega & 0 & 0 & 0 
\end{array}\right),\label{U131}\\
\hspace{-1cm}U_{2}=\left(\begin{array}{ccccc}
0 & 0 & -\bar{\alpha}\omega/k^2 & -\alpha\omega/k^2 & e^uh\omega/2 \\
0 & \partial_2u & 0 & 0 & \bar{q}\omega \\
0 & \alpha/\omega & 0 & 0 & 0 \\
0 & \bar{\alpha/\omega} & 0 & 0 & 0 \\
-2e^{-u}\bar{q}/\omega & -h/\omega & 0 & 0 & 0 
\end{array}\right),\label{U231}\\
\hspace{-1cm}U_{3}=\left(\begin{array}{ccccc}
\alpha/\omega & 0 & 0 & 0 & 0 \\
0 & \alpha/\omega & 0 & 0 & 0 \\
0 & 0 & 0 & 0 & 0 \\
0 & 0 & 0 & 0 & 0 \\
0 & 0 & 0 & 0 & 0 
\end{array}\right),\label{U331}\\
\hspace{-1cm}U_{4}=\left(\begin{array}{ccccc}
\bar{\alpha}/\omega & 0 & 0 & 0 & 0 \\
0 & \bar{\alpha}/\omega & 0 & 0 & 0 \\
0 & 0 & 0 & 0 & 0 \\
0 & 0 & 0 & 0 & 0 \\
0 & 0 & 0 & 0 & 0
\end{array}\right),\label{U431}
\end{eqnarray}
and the remaining Gauss--Codazzi equations are
\begin{eqnarray}
\partial_1\partial_2u+\frac{1}{2}e^{u}\left(h^2+\frac{4\vert\alpha\vert^2}{k^2}\right)-2e^{-u}\vert q\vert^2=0,\label{GC131}\\
\partial_2q-\frac{1}{2}e^{u}\partial_1h=0,\qquad \partial_1\bar{q}-\frac{1}{2}e^{u}\partial_2h=0,\label{GC231}
\end{eqnarray}
where 
\begin{eqnarray}
\langle\partial_1^2F,N\rangle=Q=q(x_1,x_2)\omega\in\mathbb{C},\\
\langle\partial_2^2F,N\rangle=\bar{Q}=\bar{q}(x_1,x_2)\omega\in\mathbb{C},\\
\langle\partial_1\partial_2F,N\rangle=\frac{1}{2}e^{\phi}H=\frac{1}{2}e^uh\omega,\\
H=h(x_1,x_2)/\omega\in\mathbb{R}.
\end{eqnarray}
Hence, the structural equations of case \ref{case1} of the manifold $F_3$ are linked with the immersion of a surface in a spherical space $S^3(k^2/4\vert\alpha\vert^2)$.

\subsubsection{The case \ref{case2}, when $\partial_3\phi=\partial_4\phi=0$.}~\\
As a result for the case \ref{case2}, the metric takes the form
\begin{equation}
\hspace{-1cm}g=\left(\begin{array}{cccc}
0 & \frac{1}{2}e^u & 0 & 0 \\
\frac{1}{2}e^u & 0 & 0 & 0 \\
0 & 0 & 0 & \frac{1}{2}\omega\\
0 & 0 & \frac{1}{2}\omega & 0
\end{array}\right),\qquad\begin{array}{l}
u=u(x_1,x_2)\in\mathbb{R},\\
\omega=\alpha^2\theta_3\theta_4+\vert\beta\theta_3+\bar{\beta}\theta_4\vert+\gamma^2,\\
\alpha,\gamma\in\mathbb{R},\qquad\beta\in\mathbb{C}.
\end{array}\label{gsol32}
\end{equation}
The Gauss--Weingarten equations (\ref{GW3}) are given in terms of the matrices
\begin{eqnarray}
\hspace{-1cm}U_{1}=\left(\begin{array}{ccccc}
\partial_1u & 0 & 0 & 0 & Q \\
0 & 0 & 0 &0 & e^uH/2 \\
0 & 0 & 0 & 0 & 0 \\
0 & 0 & 0 & 0 & 0 \\
-H & -2e^{-u}Q & 0 & 0 & 0 
\end{array}\right),\label{U132}\\
\hspace{-1cm}U_{2}=\left(\begin{array}{ccccc}
0 & 0 & 0 & 0 & e^uH/2 \\
0 & \partial_2u & 0 & 0 & \bar{Q} \\
0 & 0 & 0 & 0 & 0 \\
0 & 0 & 0 & 0 & 0 \\
-2e^{-u}\bar{Q} & -H & 0 & 0 & 0 
\end{array}\right),\label{U232}\\
\hspace{-1cm}U_{3}=U_4=\left(\begin{array}{ccccc}
0 & 0 & 0 & 0 & 0 \\
0 & 0 & 0 & 0 & 0 \\
0 & 0 & 0 & 0 & 0 \\
0 & 0 & 0 & 0 & 0 \\
0 & 0 & 0 & 0 & 0 
\end{array}\right),\label{U332}
\end{eqnarray}
and the remaining Gauss--Codazzi equations are
\begin{eqnarray}
\partial_1\partial_2u+\frac{1}{2}e^{u}H^2-2e^{-u}\vert Q\vert^2=0,\label{GC132}\\
\partial_2Q-\frac{1}{2}e^{u}\partial_1H=0,\qquad \partial_1\bar{Q}-\frac{1}{2}e^{u}\partial_2H=0,\label{GC232}
\end{eqnarray}
where 
\begin{eqnarray}
\langle\partial_1^2F,N\rangle=Q=Q(x_1,x_2)\in\mathbb{C},\\
\langle\partial_2^2F,N\rangle=\bar{Q}=\bar{Q}(x_1,x_2)\in\mathbb{C},\\
\langle\partial_1\partial_2F,N\rangle=\frac{1}{2}e^{u}H,\\
H=H(x_1,x_2)\in\mathbb{R}.
\end{eqnarray}
Hence, the structural equations of case \ref{case2} of the manifold $F_3$ are linked with the immersion of a surface in a Euclidean space $\mathbb{R}^3$.

\section{Structural equations of a supermanifold in a spherical$/$hyperbolic superspace}\label{SecSH}\setcounter{equation}{0}
In this section, we consider the same supermanifold $\mathcal{S}_F$, as described in section \ref{SecEuc}, but immersed in a superspace $\mathcal{M}(c)$ with a prescribed non-zero curvature such that
\begin{equation}
\hspace{-2cm}\langle F\vert F\rangle=c=c_0+c_3\xi_1\xi_2+c_2\xi_1\xi_3+c_1\xi_2\xi_3,\qquad \mathbb{R}\ni c_i\neq0,\quad i=0,1,2,3.
\end{equation}
The inner product is changed accordingly in such a way that the scalar product on each manifold $F_i$, $i=0,1,2,3$, is now the spherical ($c_i>0$) or the hyperbolic ($c_i<0$) scalar product, e.g. for a 3-dimensional space $M^3(c_i)$ with a non-zero prescribed curvature, the scalar product is
\begin{eqnarray}
\langle A,B\rangle=\mbox{sgn}(c_i)A_0B_0+A_1B_1+A_2B_2+A_3B_3,\\
A,B\in M^3(c_i)=\lbrace F\in\mathbb{C}^4\vert\langle F,F\rangle=c_i\rbrace,\qquad\mathbb{R}\ni c_i\neq0.\nonumber
\end{eqnarray}
If we consider the case where some $c_i$, $i=0,1,2,3$, are zero, than the associated structural equations of the manifolds can be obtained as in section \ref{SecEuc}. By construction, the supermanifold $\mathcal{S}_F$ satisfies the properties
\begin{equation}
\langle\partial_j\mathcal{S}_F,\mathcal{S}_F\rangle=0,\qquad j=1,2,3,4.
\end{equation}
We assume that there exists a normal vector $N$ of the form
\begin{equation}
N=N_0+N_3\xi_1\xi_2+N_2\xi_1\xi_3+N_1\xi_2\xi_3,
\end{equation}
which satisfies the properties
\begin{equation}
\hspace{-2cm}\langle N\vert N\rangle=1+\xi_1\xi_2+\xi_1\xi_3+\xi_2\xi_3,\quad \langle \mathcal{S}_F\vert N\rangle=\langle\partial_j\mathcal{S}_F\vert N\rangle=0,\quad j=1,2,3,4,
\end{equation}
i.e. such that all $N_i$, $i=0,1,2,3$, are unitary in their respective space.

Once again, we investigate each type of manifold separately, where each manifold is immersed in a hyperbolic or spherical space. One should note that for all three types of manifold, we consider the same conformal parametrisation and we drop all unnecessary quantities related to the dependency in $\theta_3$ and $\theta_4$ or to the derivatives with respect to $\theta_3$ and $\theta_4$. In addition, in the three following subsections, we will not keep the indices of the quantities which specify to which manifold ($F_0$, $F_2$ or $F_3$) they belong.

\subsection{Structural equations for a manifold of type $F_0$ in a spherical/hyperbolic space}\label{SecNEF0}
For the surface associated with $F_0$, renamed $F$ in this subsection, the metric takes the form
\begin{equation}
\hspace{-2cm}g=[g_{ij}]=\left(\begin{array}{cc}
0 & \frac{1}{2}e^{u} \\
\frac{1}{2}e^{u} & 0
\end{array}\right),\qquad \begin{array}{l}
g_{ij}=\langle \partial_iF,\partial_jF\rangle,\qquad i,j=1,2,\\
\langle F,F\rangle=c,
\end{array}\label{NEg0}
\end{equation}
where $u$ is a real function of $x_1$ and $x_2$. We obtain the well-known geometry of a conformally parametrised surface immersed in a 3-dimensional hyperbolic or spherical space. The moving frame on the surface is given by
\begin{equation}
\Omega=\left(\begin{array}{cccc}
\partial_1F, & \partial_2F, & N, & F
\end{array}\right)^T.\label{NErep0}
\end{equation}
Assuming that we can write the second derivatives of the immersion function $F$ and the first derivatives of the normal unit vector $N$ as a linear combination of the elements of the moving frame $\Omega$, i.e.
\begin{eqnarray}
\partial_i\partial_jF&=\sum_{k=1}^2\Gamma_{ij}^k\partial_kF+b_{ij}N+\kappa_{ij}F,&\qquad i,j=1,2,\label{NEDDF0}\\
\hspace{4mm}\partial_iN&=\sum_{j=1}^{2}-b_{i}^j\partial_jF+\omega_{i}N+\mu_iF,&\qquad i=1,2,\label{NEDN0}
\end{eqnarray}
we can construct the Gauss--Weingarten equations in the form
\begin{equation}
\partial_i\Omega=U_{i}\Omega,\qquad i=1,2,\label{NEGW0}
\end{equation}
where
\begin{equation}
U_{i}=\left(\begin{array}{cccc}\label{NEUi0}
\Gamma_{i1}^1 & \Gamma_{i1}^2 & b_{i1} & \kappa_{i1}\\
\Gamma_{i2}^1 & \Gamma_{i2}^2 & b_{i2} & \kappa_{i2}\\
-b_{i}^1 & -b_{i}^2 & \omega_{i} & \mu_i \\
\delta_{i1} & \delta_{i2} & 0 & 0
\end{array}\right),
\end{equation}
and $\delta_{ij}$, $i,j=1,2$, is the Kronecker delta function. In addition, the Gauss--Codazzi equations are obtained through the compatibility condition of the Gauss--Weingarten equations (\ref{NEGW0}), i.e.
\begin{equation}
\partial_1U_{2}-\partial_2U_{1}+[U_{2},U_{1}]=0.\label{NEZCC0}
\end{equation}
Explicitly, the Gauss--Weingarten equations (\ref{NEGW0}) are given in terms of the matrices
\begin{eqnarray}
U_{1}=\left(\begin{array}{cccc}
\partial_1 u & 0 & Q & 0 \\
0 & 0 & \frac{1}{2}e^uH & -e^{u}/2c \\
-H & -2e^{-u}Q & 0 & 0 \\
1 & 0 & 0 & 0
\end{array}\right),\label{NEU10}\\
U_{2}=\left(\begin{array}{cccc}
0 & 0 & \frac{1}{2}e^uH & -e^u/2c \\
0 & \partial_2u & \bar{Q} & 0 \\
-2e^{-u}\bar{Q} & -H & 0 & 0 \\
0 & 1 & 0 & 0
\end{array}\right),\label{NEU20}
\end{eqnarray}
and the Gauss--Codazzi equations are
\begin{eqnarray}
\partial_1\partial_2u+\frac{1}{2}e^{u}\left(H^2+\frac{1}{c}\right)-2e^{-u}\vert Q\vert^2=0,\label{NEGC10}\\
\partial_2Q-\frac{1}{2}e^{u}\partial_1H=0,\qquad \partial_1\bar{Q}-\frac{1}{2}e^{u}\partial_2H=0,\label{NEGC20}
\end{eqnarray}
where $Q=Q(x_1,x_2)=\langle \partial_1^2F,N\rangle\in\mathbb{C}$ is associated with the Hopf differential and $H=H(x_1,x_2)=2e^{-u}\langle \partial_1\partial_2F,N\rangle\in\mathbb{R}$ is the mean curvature of the surface.

\subsection{Structural equations for a manifold of type $F_2$ in a spherical/hyperbolic space}\label{SecNEF2}
For the manifold associated with $F_2$, renamed $F$ in this subsection, the metric tensor takes the form
\begin{equation}
\hspace{-2cm}g=[g_{ij}]=\left(\begin{array}{ccc}
0 & \frac{1}{2}e^{\phi} & 0 \\
\frac{1}{2}e^{\phi} & 0 & 0 \\
0 & 0 & e^\psi
\end{array}\right),\qquad\begin{array}{l}
g_{ij}=\langle \partial_iF,\partial_jF\rangle,\qquad i,j=1,2,\\
\langle F,F\rangle=c,
\end{array}\label{NEg2}
\end{equation}
where $\psi$ is a real function of $x_1$ and $x_2$ and $\phi$ is a real function of $x_1$, $x_2$ and $\theta_3$ satisfying $\partial_3^3e^\phi=0$. The moving frame on the manifold is given by
\begin{equation}
\Omega=\left(\begin{array}{ccccc}
\partial_1F, & \partial_2F, & \partial_3F, & N, & F
\end{array}\right)^T.\label{NErep2}
\end{equation}
Assuming that we can write the second derivatives of the immersion function $F$ and the first derivatives of the normal unit vector $N$ as a linear combination the elements of the moving frame $\Omega$, i.e.
\begin{eqnarray}
\partial_i\partial_jF&=\sum_{k=1}^3\Gamma_{ij}^k\partial_kF+b_{ij}N+\kappa_{ij}F,&\qquad i,j=1,2,3,\label{NEDDF2}\\
\hspace{4mm}\partial_iN&=\sum_{j=1}^{3}-b_{i}^j\partial_jF+\omega_{i}N+\mu_iF,&\qquad i=1,2,3,\label{NEDN2}
\end{eqnarray}
we obtain that the manifold $F$ does not depend on $\theta_3$. Hence, the structural equations of the manifold $F_2$ immersed in a spherical/hyperbolic space are a copy of the structural equations of the surface $F_0$ immersed in a spherical/hyperbolic space.

\subsection{Structural equations for a manifold of type $F_3$ in a spherical/hyperbolic space}\label{SecNEF3}
For the manifold associated with $F_3$, renamed $F$ in this subsection, the metric takes the form
\begin{equation}
\hspace{-2cm}g=[g_{ij}]=\left(\begin{array}{cccc}
0 & \frac{1}{2}e^{\phi} & 0 & 0 \\
\frac{1}{2}e^{\phi} & 0 & 0 & 0 \\
0 & 0 & 0 & \frac{1}{2}e^{\psi} \\
0 & 0 & \frac{1}{2}e^{\psi} & 0
\end{array}\right),\qquad\begin{array}{l}
g_{ij}=\langle \partial_iF,\partial_jF\rangle,\qquad i,j=1,2,\\
\langle F,F\rangle=c,
\end{array}\label{NEg3}
\end{equation}
where $\psi$ and $\phi$ are real functions of $x_1$, $x_2$ $\theta_3$ and $\theta_4$ satisfying $\partial_3^3e^\phi=\partial_4^3e^\phi=0$ and $\partial_3^2e^\psi=\partial_4^2e^\psi=0$. The moving frame on the manifold is given by
\begin{equation}
\Omega=\left(\begin{array}{cccccc}
\partial_1F, & \partial_2F, & \partial_3F, & \partial_4F, & N, & F
\end{array}\right)^T.\label{NErep3}
\end{equation}
Assuming that we can write the second derivatives of the immersion function $F$ and the first derivatives of the normal unit vector $N$ as a linear combination of the elements of the moving frame $\Omega$, i.e.
\begin{eqnarray}
\partial_i\partial_jF&=\sum_{k=1}^4\Gamma_{ij}^k\partial_kF+b_{ij}N+\kappa_{ij}F,&\qquad i,j=1,2,3,4,\label{NEDDF3}\\
\hspace{4mm}\partial_iN&=\sum_{j=1}^{4}-b_{i}^j\partial_jF+\omega_{i}N+\mu_iF,&\qquad i=1,2,3,4,\label{NEDN3}
\end{eqnarray}
we can construct the Gauss--Weingarten equations in the form
\begin{equation}
\partial_i\Omega=U_{i}\Omega,\qquad i=1,2,3,4,\label{NEGW3}
\end{equation}
where
\begin{equation}
U_{i}=\left(\begin{array}{cccccc}\label{NEUi3}
\Gamma_{i1}^1 & \Gamma_{i1}^2 & \Gamma_{i1}^3 & \Gamma_{i1}^4 & b_{i1} & \kappa_{i1} \\
\Gamma_{i2}^1 & \Gamma_{i2}^2 & \Gamma_{i2}^3 & \Gamma_{i2}^4 & b_{i2} & \kappa_{i2} \\
\Gamma_{i3}^1 & \Gamma_{i3}^2 & \Gamma_{i3}^3 & \Gamma_{i3}^4 & b_{i3} & \kappa_{i3} \\
\Gamma_{i4}^1 & \Gamma_{i4}^2 & \Gamma_{i4}^3 & \Gamma_{i4}^4 & b_{i4} & \kappa_{i4} \\
-b_{i}^1 & -b_{i}^2 & -b_{i}^3 & -b_{i}^4 & \omega_{i} & \mu_i \\
\delta_{i1} & \delta_{i2} & \delta_{i3} & \delta_{i4} & 0 & 0 
\end{array}\right)
\end{equation}
and $\delta_{ij}$ is the Kronecker delta function. In addition, the Gauss--Codazzi equations are obtained through the compatibility condition of the Gauss--Weingarten equations (\ref{NEGW3}), i.e.
\begin{equation}
\partial_iU_{j}-\partial_jU_{i}+[U_{j},U_{i}]=0,\qquad i,j=1,2,3,4.\label{NEZCC3}
\end{equation}
As a result, we get a set of PDEs from which we get that the constant $c$ must be negative ($c<0$) and that the solutions split into two cases, namely:
\begin{enumerate}
\item $\partial_1\psi=\partial_2\psi=0$,\label{NEcase1}
\item $\partial_3\phi=\partial_4\phi=0$.\label{NEcase2}
\end{enumerate}
The explicit set of PDEs can be found in Appendix B.

\subsubsection{The case \ref{NEcase1}, when $\partial_1\psi=\partial_2\psi=0$.}~\\
As a result for the case \ref{NEcase1}, the metric takes the form
\begin{equation}
\hspace{-1cm}g=\left(\begin{array}{cccc}
0 & \frac{1}{2}e^u\omega^2 & 0 & 0 \\
\frac{1}{2}e^u\omega^2 & 0 & 0 & 0 \\
0 & 0 & 0 & \frac{1}{2}k^2\\
0 & 0 & \frac{1}{2}k^2 & 0
\end{array}\right),\qquad\begin{array}{l}
u=u(x_1,x_2)\in\mathbb{R},\\
\omega=\alpha\theta_3\theta_4+\beta\theta_3+\bar{\beta}\theta_4+\gamma,\\
k,\alpha,\gamma\in\mathbb{R},\qquad\beta\in\mathbb{C}.
\end{array}\label{NEgsol31}
\end{equation}
The Gauss--Weingarten equations (\ref{NEGW3}) are given by the matrices
\begin{eqnarray}
\hspace{-2cm}U_{1}=\left(\begin{array}{cccccc}
\partial_1u & 0 & 0 & 0 & 0 & 0 \\
0 & 0 & -e^u\omega\partial_4\omega/k^2 & -e^u\omega\partial_3\omega/k^2 & \frac{1}{2}e^uH\omega^2 & -e^u\omega^2/2c \\
\partial_3\omega/\omega & 0 & 0 & 0 & 0 & 0 \\
\partial_4\omega/\omega & 0 & 0 & 0 & 0 & 0 \\
-H & 0 & 0 & 0 & 0 & 0 \\
1 & 0 & 0 & 0 & 0 & 0
\end{array}\right),\label{NEU131}\\
\hspace{-2cm}U_{2}=\left(\begin{array}{cccccc}
0 & 0 & -e^u\omega\partial_4\omega/k^2 & -e^u\omega\partial_3\omega/k^2 & \frac{1}{2}e^uH\omega^2 & -e^u\omega^2/2c \\
0 & \partial_2u & 0 & 0 & 0 & 0 \\
0 & \partial_3\omega/\omega & 0 & 0 & 0 & 0 \\
0 & \partial_4\omega/\omega & 0 & 0 & 0 & 0 \\
0 & -H & 0 & 0 & 0 & 0 \\
0 & 1 & 0 & 0 & 0 & 0
\end{array}\right),\label{NEU231}\\
\hspace{-2cm}U_{3}=\left(\begin{array}{cccccc}
\partial_3\omega/\omega & 0 & 0 & 0 & 0 & 0 \\
0 & \partial_3\omega/\omega & 0 & 0 & 0 & 0 \\
0 & 0 & 0 & 0 & 0 & 0 \\
0 & 0 & 0 & 0 & G/2k^2 & -1/2ck^2 \\
0 & 0 & -G & 0 & 0 & 0 \\
0 & 0 & 1 & 0 & 0 & 0
\end{array}\right),\label{NEU331}\\
\hspace{-2cm}U_{4}=\left(\begin{array}{cccccc}
\partial_4\omega/\omega & 0 & 0 & 0 & 0 & 0 \\
0 & \partial_4\omega/\omega & 0 & 0 & 0 & 0 \\
0 & 0 & 0 & 0 & G/2k^2 & -1/2ck^2 \\
0 & 0 & 0 & 0 & 0 & 0 \\
0 & 0 & 0 & -G & 0 & 0 \\
0 & 0 & 0 & 1 & 0 & 0
\end{array}\right),\label{NEU431}
\end{eqnarray}
where the solution is
\begin{eqnarray}
\hspace{-1cm}\langle\partial_1\partial_2F,N\rangle=\frac{1}{2}e^{\phi}H=\frac{1}{2}e^u\omega^2H,\qquad H=-\epsilon\sqrt{-c}\left(\frac{1}{c}+\frac{2\alpha}{k^2\omega}\right)\in\mathbb{R},\\
\hspace{-1cm}\langle\partial_3\partial_4F,N\rangle=\frac{1}{2}e^{\psi}G=\frac{1}{2}k^2G,\qquad G=\epsilon\sqrt{-c}\in\mathbb{R},\qquad \epsilon^2=1,\\
\hspace{-1cm}u=\phi-\ln(\omega^2)=\ln\left(\frac{2\vert\partial_1f\vert^2}{\sigma^2(1+\vert f\vert^2)^2}\right),\qquad f=f(x_1),\quad\bar{f}=\bar{f}(x_2)\in\mathbb{C}.
\end{eqnarray}
The structural equations of the case \ref{NEcase1} of the manifold $F_3$ are linked with solutions of the Liouville equation
\begin{equation}
\hspace{-2cm}\partial_1\partial_2u=\frac{2}{k^2}\left(\alpha\gamma+\frac{c\alpha^2}{k^2}-\vert\beta\vert^2\right)e^u=-2\sigma^2 e^u,\qquad \alpha\gamma<\vert\beta\vert^2-\frac{c\alpha^2}{k^2},\quad\sigma\in\mathbb{R}.
\end{equation}
The geometric quantities (i.e. the functions $H$, $G$, $\phi$ and $\psi$) are determined.
~\\
\subsubsection{The case \ref{NEcase2}, $\partial_3\phi=\partial_4\phi=0$.}~\\
As a result for the case \ref{NEcase2}, the metric takes the form
\begin{equation}
\hspace{-1cm}g=\left(\begin{array}{cccc}
0 & \frac{1}{2}e^\phi & 0 & 0 \\
\frac{1}{2}e^\phi & 0 & 0 & 0 \\
0 & 0 & 0 & \frac{1}{2}e^\psi\\
0 & 0 & \frac{1}{2}e^\psi & 0
\end{array}\right),\qquad\begin{array}{l}
\phi=\phi(x_1,x_2)\in\mathbb{R},\\
\psi=\psi(x_1,x_2)\in\mathbb{R}.
\end{array}\label{NEgsol32}
\end{equation}
The Gauss--Weingarten equations (\ref{NEGW3}) are given by the matrices
\begin{eqnarray}
\hspace{-1cm}U_{1}=\left(\begin{array}{cccccc}
\partial_1\phi & 0 & 0 & 0 & Q & 0 \\
0 & 0 & 0 & 0 & \frac{1}{2}e^\phi H & -e^\phi/2c \\
0 & 0 & \partial_1\psi/2 & 0 & 0 & 0 \\
0 & 0 & 0 & \partial_1\psi/2 & 0 & 0 \\
-H & -2e^{-\phi}Q & 0 & 0 & 0 & 0 \\
1 & 0 & 0 & 0 & 0 & 0
\end{array}\right),\label{NEU132}\\
\hspace{-1cm}U_{2}=\left(\begin{array}{cccccc}
0 & 0 & 0 & 0 & \frac{1}{2}e^\phi H & -e^\phi/2c \\
0 & \partial_2\phi & 0 & 0 & \bar{Q} & 0 \\
0 & 0 & \partial_2\psi/2 & 0 & 0 & 0 \\
0 & 0 & 0 & \partial_2\psi/2 & 0 & 0 \\
-2e^{-\phi}\bar{Q} & -H & 0 & 0 & 0 & 0 \\
0 & 1 & 0 & 0 & 0 & 0
\end{array}\right),\label{NEU232}\\
\hspace{-1cm}U_{3}=\left(\begin{array}{cccccc}
0 & 0 & \partial_1\psi/2 & 0 & 0 & 0 \\
0 & 0 & \partial_2\psi/2 & 0 & 0 & 0 \\
0 & 0 & 0 & 0 & 0 & 0 \\
-e^{\psi-\phi}\partial_2\psi/2 & -e^{\psi-\phi}\partial_1\psi/2 & 0 & 0 & \frac{1}{2}e^\psi G & -e^\psi/2c \\
0 & 0 & -G & 0 & 0 & 0 \\
0 & 0 & 1 & 0 & 0 & 0
\end{array}\right),\label{NEU332}\\
\hspace{-1cm}U_4=\left(\begin{array}{cccccc}
0 & 0 & 0 & \partial_1\psi/2 & 0 & 0 \\
0 & 0 & 0 & \partial_2\psi/2 & 0 & 0 \\
-e^{\psi-\phi}\partial_2\psi/2 & -e^{\psi-\phi}\partial_1\psi/2 & 0 & 0 & \frac{1}{2}e^\psi G & -e^\psi/2c \\
0 & 0 & 0 & 0 & 0 & 0 \\
0 & 0 & 0 & -G & 0 & 0 \\
0 & 0 & 0 & 1 & 0 & 0
\end{array}\right),\label{NEU432}
\end{eqnarray}
and the remaining Gauss--Codazzi equations are reduced to one equation
\begin{eqnarray}
\hspace{-2cm}\partial_1\partial_2\phi=2e^{-\phi}\vert Q\vert^2-\frac{1}{2}e^\phi\left(H^2+\frac{1}{c}\right)\\
\hspace{-2cm}\phantom{\partial_1\partial_2\phi}=\frac{e^\phi}{c}+\frac{\partial_1\partial_2\psi}{1+ce^{-\phi}\partial_1\psi\partial_2\psi}+\frac{ce^{-\phi}/2}{1+ce^{-\phi}\partial_1\psi\partial_2\psi}\left(\left(\partial_1\partial_2\psi+\frac{\partial_1\psi\partial_2\psi}{2}\right)^2\right.\nonumber\\
\left.+(2\partial_1\phi\partial_1\psi-2\partial_1^2\psi-(\partial_1\psi)^2)(2\partial_2\phi\partial_2\psi-2\partial_2^2\psi-(\partial_2\psi)^2)\right)\label{NEGC232}
\end{eqnarray}
subject to the constraint
\begin{equation}
\partial_1\psi\partial_2\psi<-\frac{e^\phi}{c}.\label{NEGC132}
\end{equation}
The functions $Q$, $\bar{Q}$, $H$ and $G$ can be written explicitly in terms of $\phi$ and $\psi$, i.e.
\begin{eqnarray}
\langle\partial_1^2F,N\rangle=Q=Q(x_1,x_2)\in\mathbb{C},\\
\langle\partial_2^2F,N\rangle=\bar{Q}=\bar{Q}(x_1,x_2)\in\mathbb{C},\\
\langle\partial_1\partial_2F,N\rangle=\frac{1}{2}e^{u}H\in\mathbb{R},\\
\langle\partial_3\partial_4F,N\rangle=\frac{1}{2}e^{\psi}G\in\mathbb{R},\\
Q=\epsilon\frac{\partial_1\phi\partial_1\psi-\partial_1^2\psi-\frac{1}{2}(\partial_1\psi)^2}{\sqrt{-c^{-1}-e^{-\phi}\partial_1\psi\partial_2\psi}},\qquad\epsilon^2=1,\\
\bar{Q}=\epsilon\frac{\partial_2\phi\partial_2\psi-\partial_2^2\psi-\frac{1}{2}(\partial_2\psi)^2}{\sqrt{-c^{-1}-e^{-\phi}\partial_1\psi\partial_2\psi}},\\
H=-\epsilon\frac{c^{-1}+e^{-\phi}(\partial_1\partial_2\psi+\frac{1}{2}\partial_1\psi\partial_2\psi)}{\sqrt{-c^{-1}-e^{-\phi}\partial_1\psi\partial_2\psi}},\\
G=\epsilon\sqrt{\frac{-1}{c}-e^{-\phi}\partial_1\psi\partial_2\psi}.
\end{eqnarray}

\section{Conclusions}\label{SecConc}\setcounter{equation}{0}
In this paper, we have investigated the structural equations of supermanifolds with two bosonic and two fermionic independent variables through a moving frame formalism. We considered supermanifolds immersed in a Euclidean superspace, in a hyperbolic superspace and in a spherical superspace. By splitting the supermanifold into manifolds for each Grassmann number component, we obtained that the supermanifold can be expressed as copies of three different types of manifold in the Euclidean case and copies of two different types of manifold in the spherical and hyperbolical cases.

For the Euclidean superspace, considering the body-like type of manifold ($F_0$), we obtain the same structural equations as for a surface immersed in a Euclidean space, i.e. the classical Gauss--Weingarten and the Gauss--Codazzi equations. This result ensures that, when we take the limit where all fermionic quantities vanish, we obtain the classical (non-super) differential geometry. For the manifold type called $F_2$ in a Euclidean space, we obtain that the structural equations are linked with the structural equations of a surface immersed in a spherical space (not Euclidean). For the manifold type called $F_3$ immersed in a Euclidean space, the solutions to the structural equations split into two non-trivial cases. For the case \ref{case1}, where the metric coefficient in the direction of the bosonic variables (i.e. $g_{12}$) depends on the bosonic and fermionic variables and the metric coefficient in the direction of the fermionic variables (i.e. $g_{34}$) is constant, we obtain once again that the solution of the structural equations is linked with those of a surface immersed in a spherical space. For the case \ref{case2}, where the metric coefficient in the direction of the bosonic variables (i.e. $g_{12}$) depends only on the bosonic variables and the metric coefficent in the direction of the fermionic variables (i.e. $g_{34}$) depends only on the fermionic variables, we obtain that the solution of the structural equations is linked to those of a surface immersed in a Euclidean space.

In the hyperbolic and the spherical cases, we obtain that the manifold called $F_2$ does not depend on the fermionic variables. Hence, it is a copy of the body-like manifold (called $F_0$). The body-like manifold $F_0$ is equivalent to a surface immersed in a hyperbolic or spherical space. This result ensures that, when we take the limit where all fermionic quantities vanish, we obtain the classical (non-super) differential geometry. For the manifold type called $F_3$, we obtain that the immersion can only exist in a hyperbolic space. In addition, a new function (called $G$) is needed to construct the structural equations. Once again, these structural equations split into two non-trivial cases. For the case \ref{NEcase1}, where the metric coefficient in the direction of the bosonic variables (i.e. $g_{12}$) depends on all independent variables and the metric coefficient in the direction of the fermionic variables (i.e. $g_{34}$) is constant, we were able to completly solve the structural equations. Also, it should be noted that the quantity $G$ is constant and the metric is linked with the Liouville equation, which is related to the supersymmetric minimal surface equations \cite{GH17}. In the classical case, this development leads to a new analytic and geometric approach for minimal surfaces via orthogonal polynomials \cite{DG15}. The question arises as to whether such relations can also be established in the supersymetric case. For the case \ref{NEcase2}, where the metric does not depend on the fermionic variables, we were able to express all geometric quantities (associated with the Hopf differential $Q$, $\bar{Q}$, the mean curvature $H$ and the new quantity $G$) in terms of the metric coefficents. There remains a non-trivial link between the two metric coefficients provided by a non-linear PDE and a differential inequality. The immersion into a spherical and hyperbolic superspace is more restrictive than in a Euclidean superspace due to the fact that the geometric quantities are completely defined in the first case for non-body-like manifolds compared to the addition of a parameter in the Euclidean superspace.

This investigation can be extended in many directions. It would be interesting to investigate the immersion of supermanifolds in other types of superspaces to check if the structural equations allow more or less freedom than for the Euclidean superspace. It would also be interesting to consider a higher number of fermionic variables to check if one gets the same three or two types of manifolds or if one gets new types of manifolds. Since we obtained a new function $G$ in the structural equations for immersions in a hyperbolic superspace, it would be interesting to investigate the geometric meaning of this function and whether such a geometric characterization is complete up to certain affine transformations. A classification of the supermanifold could also be undertaken. It would be interesting to investigate the properties of the supermanifold if we consider solutions linked with minimal surfaces, especially since we obtained the Liouville equation in one of the cases of the hyperbolic superspace. In the classical (non-super) geometry of surfaces, Bonnet surfaces are known for their solutions linked with the Painlev\'e P6 equation (see e.g. \cite{Conte17} and references therein). It would be interesting to investigate if similar properties appear in a Bonnet-like supermanifold.

\section*{Acknowledgements}
SB was partially supported by a doctoral fellowship provided by the Facult\'e des \'Etudes Sup\'erieures et Postdoctorales of the Universit\'e de Montr\'eal and by a postdoctoral fellowship provided by the Fonds de Recherche du Qu\'ebec : Nature et Technologie (FQRNT). AMG was supported by a research grant provided by NSERC of Canada.

\section*{Appendix A}
The (previously unsolved) structural equations of section \ref{SecF3} are
\begin{eqnarray*}
\partial_1\partial_2\phi+\frac{1}{2}e^\phi H^2-2e^{-\phi}\vert Q\vert^2+\frac{1}{2}e^{\phi-\psi}\partial_3\phi\partial_4\phi=0,\\
\partial_2Q-\frac{1}{2}e^\phi\partial_1H=0,\qquad \partial_1\bar{Q}-\frac{1}{2}e^\phi\partial_2H=0,\\
\partial_1\partial_3\phi=\partial_1\partial_4\phi=\partial_2\partial_3\phi=\partial_2\partial_4\phi=0,\\
\partial_1\psi\partial_2\psi=\partial_3\phi\partial_3\psi=\partial_3\phi\partial_4\psi=\partial_4\phi\partial_3\psi=\partial_4\phi\partial_4\psi=0,\\
2\partial_3^2\phi+(\partial_3\phi)^2=0,\qquad 2\partial_4^2\phi+(\partial_4\phi)^2=0,\\
2\partial_3\partial_4\phi+\partial_3\phi\partial_4\phi=0,\\
\partial_3Q-\frac{1}{2}Q\partial_3\phi=0,\qquad\partial_4Q-\frac{1}{2}Q\partial_4\phi=0,\\
\partial_3H+\frac{1}{2}H\partial_3\phi=0,\qquad\partial_4H+\frac{1}{2}H\partial_4\phi=0,\\
\partial_3\bar{Q}-\frac{1}{2}\bar{Q}\partial_3\phi=0,\qquad\partial_4\bar{Q}-\frac{1}{2}\bar{Q}\partial_4\phi=0,\\
\partial_3^3e^\phi=\partial_4^3e^\phi=0,\qquad \partial_3^2e^\psi=\partial_4^2e^\psi=0.
\end{eqnarray*}

\section*{Appendix B}
The (previously unsolved) structural equations of section \ref{SecNEF3} are
\begin{eqnarray*}
\hspace{-2cm}\partial_1\partial_2\phi+\frac{1}{2}\left(H^2+\frac{1}{c}\right)-2e^{-\phi}\vert Q\vert^2+\frac{1}{2}e^{\phi-\psi}\partial_3\phi\partial_4\phi=0,\\
\hspace{-2cm}\partial_2Q-\frac{1}{2}e^\phi\partial_1H=0,\qquad\partial_1\bar{Q}-\frac{1}{2}e^\phi\partial_2H=0,\\
\hspace{-2cm}\partial_3Q-\frac{1}{2}Q\partial_3\phi,\qquad\partial_3H+\frac{1}{2}(H-G)\partial_3\phi=0,\qquad\partial_3\bar{Q}-\frac{1}{2}\bar{Q}\partial_3\phi=0,\\
\hspace{-2cm}\partial_4Q-\frac{1}{2}Q\partial_4\phi,\qquad\partial_4H+\frac{1}{2}(H-G)\partial_4\phi=0,\qquad\partial_4\bar{Q}-\frac{1}{2}\bar{Q}\partial_4\phi=0,\\
\hspace{-2cm}\partial_1G+\frac{1}{2}(G-H)\partial_1\psi-e^{-\phi}Q\partial_2\psi=0,\qquad\partial_2G+\frac{1}{2}(G-H)\partial_2\psi-e^{-\phi}\bar{Q}\partial_1\psi=0,\\
\hspace{-2cm}\frac{1}{c}+G^2+e^{-\phi}\partial_1\psi\partial_2\psi=0,\qquad \partial_3\psi=\partial_4\psi=0,\qquad \partial_3G=\partial_4G=0,\\
\hspace{-2cm}\frac{1}{c}+GH+e^{-\psi}\left(\partial_3\partial_4\phi+\frac{1}{2}\partial_3\phi\partial_4\phi\right)+e^{-\phi}\left(\partial_1\partial_2\psi+\frac{1}{2}\partial_1\psi\partial_2\psi\right)=0,\\
\hspace{-2cm}\partial_1\psi\partial_3\phi=\partial_1\psi\partial_4\phi=\partial_2\psi\partial_3\phi=\partial_2\psi\partial_4\phi=0,\\
\hspace{-2cm}\partial_1\partial_3\phi=\partial_1\partial_4\phi=\partial_2\partial_3\phi=\partial_2\partial_4\phi=0,\\
\hspace{-2cm}2\partial_1^2\psi+(\partial_1\psi)^2-2\partial_1\phi\partial_1\psi+4GQ=0,\qquad 2\partial_2^2\psi+(\partial_2\psi)^2-2\partial_2\phi\partial_2\psi+4G\bar{Q}=0,\\
\hspace{-2cm}2\partial_3^2\phi+(\partial_3\phi)^2=0,\qquad2\partial_4^2\phi+(\partial_4\phi)^2=0,\qquad\partial_3^3e^\phi=\partial_4^3e^\phi=0.
\end{eqnarray*}


\section*{References}
\begin{thebibliography}{10}  

\bibitem{BE00}
Bobenko AI and Eitner U (2000) Painlev\'e equations in the differential geometry of surfaces, Lecture notes in mathematics 1753 (Springer, New York), DOI:\href{http://dx.doi.org/10.1007/b76883}{10.1007/b76883}

\bibitem{Cartan53}
Cartan E (1904) Sur la structure des groupes infinis,  Ann. Ec. Nor. \textbf{3} XXI ,153--206.

\bibitem{Conte17}
Conte R (2017) Generalized Bonnet surfaces and Lax pairs of $P_{VI}$, J. Math. Phys. \textbf{58} 103508, DOI:\href{http://dx.doi.org/10.1063/1.4995689}{10.1063/1.4995689}

\bibitem{Davydov79}
Davydov AS (1979) Solitons in Molecular Systems, Phys. Scr. \textbf{20} 387--394, DOI:\href{http://dx.doi.org/10.1088/0031-8949/20/3-4/013}{10.1088/0031-8949/20/3-4/013}

\bibitem{DG15}
Doliwa A and Grundland AM (2015) Minimal surfaces in the soliton surface approach, arXiv:\href{http://arxiv.org/abs/1511.02173}{1511.02173}

\bibitem{Enneper}
Enneper A (1868) Analytisch-geometrische Untersuchungen, K\'onigl. Gesell. Wissensch. Georg-Augustus-Univ. Gottigen \textbf{12} 258

\bibitem{FG96}
Fokas AS and Gel'fand IM 1996 Surfaces on Lie groups, on Lie algebras, and their integrability, Comm. Math. Phys. \textbf{177} 203--220, DOI:\href{http://dx.doi.org/10.1007/BF02102436}{10.1007/BF02102436}

\bibitem{FGFL}
Fokas AS, Gel'fand IM, Finkel F and Liu QM (2000) A formula for constructing infinitely many surfaces on Lie algebras and integrable equations, Sel. Math. \textbf{6} 347--375, DOI:\href{http://dx.doi.org/10.1007/PL00001392}{10.1007/PL00001392}

\bibitem{GG10}
Goldstein PP and Grundland AM (2010) Invariant recurrence relations for $\mathbb{C}P^{N-1}$ models, J. Phys. A: Math. Theor. 43 265206 (18pp), DOI:\href{http://dx.doi.org/10.1088/1751-8113/43/26/265206}{10.1088/1751-8113/43/26/265206}

\bibitem{GGP12}
Goldstein G, Grundland AM and Post S (2012) Soliton surfaces associated with sigma models: differential and algebraic aspects, J. Phys. A: Math. Theor. \textbf{45} 395208 (19pp), DOI:\href{http://dx.doi.org/10.1088/1751-8113/45/39/395208}{10.1088/1751-8113/45/39/395208}

\bibitem{GPW91}
Gross DJ, Piran T and Weinberg S (1991) Two Dimensional Quantum Gravity and Random Surfaces, World Scientific, Singapore

\bibitem{GH17}
Grundland AM and Hariton A (2017) Algebraic aspects of the supersymmetric minimal surface equation, Symmetry \textbf{9} 318 (19pp), DOI:\href{http://dx.doi.org/10.3390/sym9120318}{10.3390/sym9120318}

\bibitem{GP11}
Grundland AM and Post S 2011 Soliton surfaces associated with generalized symmetries of integrable equations, J. Phys. A: Math. Theor. \textbf{44} 165203 (31pp), DOI:\href{http://dx.doi.org/10.1088/1751-8113/44/16/165203}{10.1088/1751-8113/44/16/165203}

\bibitem{Guest97}
Guest MA (1997) \textit{Harmonic Maps, Loop Groups, and Integrable Systems}, Cambridge University Press, Cambridge

\bibitem{Helein01}
H\'elein F (2001) \textit{Constant Mean Curvature Surfaces, Harmonic Maps and Integrable Systems}, Birkhäuser Basel, Boston, DOI:\href{http://dx.doi.org/10.1007/978-3-0348-8330-6}{10.1007/978-3-0348-8330-6}

\bibitem{JMN14}
Jensen GR, Musso E and Nicolodi L (2104) The geometric Cauchy problem for the membrane shape equation, J. Phys. A: Math. Theor. \textbf{47} 495201 (22pp), DOI:\href{http://dx.doi.org/10.1088/1751-8113/47/49/495201}{10.1088/1751-8113/47/49/495201}

\bibitem{Konopelchenko96}
Konopelchenko BG (1996) Induced Surfaces and Their Integrable Dynamics, Stud. Appl. Math. \textbf{96} 9–-51, DOI:\href{http://dx.doi.org/10.1002/sapm19969619}{10.1002/sapm19969619}

\bibitem{Landolfi03}
Landolfi G (2003) New results on the Canham--Helfrich membrane model via the generalized Weierstrass representation, J. Phys. A: Math. Gen. \textbf{36} 11937–-11954, DOI:\href{http://dx.doi.org/10.1088/0305-4470/36/48/003}{10.1088/0305-4470/36/48/003}

\bibitem{MS04}
Manton N and Sutcliffe P (2004) \textit{Topological Solitons}, Cambridge University Press, DOI:\href{http://dx.doi.org/10.1017/CBO9780511617034}{10.1017/CBO9780511617034}

\bibitem{Mikhailov86}
Mikhailov AV (1986) Integrable Magnetic Models, in \textit{Solitons}, Mod. Prob. Cond. Mat. Sc. \textbf{17} 623--690, DOI:\href{http://dx.doi.org/10.1016/B978-0-444-87002-5.50019-9}{10.1016/B978-0-444-87002-5.50019-9}

\bibitem{NS94}
Nomizu K and Sasaki T (1994) \textit{Affine Differential Geometry: Geometry of Affine Immersions}, Cambridge University Press, Cambridge

\bibitem{PS91}
Polchinski J and Strominger A (1991) Effective string theory Phys. Rev. Lett. 67, 1681--1684, DOI:\href{http://dx.doi.org/10.1103/PhysRevLett.67.1681}{10.1103/PhysRevLett.67.1681}

\bibitem{Rajaraman02}
Rajaraman R (2002) Solitons in quantum Hall systems, Eur. Phys. J. B  \textbf{29} 157--162, DOI:\href{http://dx.doi.org/10.1140/epjb/e2002-00277-7}{10.1140/epjb/e2002-00277-7}

\bibitem{RS02}
Rogers C and Schief W K 2002 \textit{B\"acklund and Darboux Transformations : Geometry and Modern Applications in Soliton Theory}, Cambridge University Press, Cambridge, DOI:\href{http://dx.doi.org/10.1017/CBO9780511606359}{10.1017/CBO9780511606359}

\bibitem{Safran94}
Safran SA (1994) \textit{Statistical Thermodynamics of Surfaces, Interfaces, and Membranes}, Addison--Wesley, Massachusetts

\bibitem{Sym83}
Sym A (1984) Soliton surfaces, Lett. Nuovo Cimento \textbf{41} 33--40, DOI:\href{http://dx.doi.org/10.1007/BF02748459}{10.1007/BF02748459}

\bibitem{Uhlenbeck89}
Uhlenbeck K (1989) Harmonic maps into Lie groups: classical solutions of the chiral model, J. Diff. Geom. \textbf{30} 1--50, DOI:\href{http://dx.doi.org/10.4310/jdg/1214443286}{10.4310/jdg/1214443286}

\bibitem{Weierstrass}
Weierstrass K (1866) Fortsetzung der Untersuchung \"uber die Minimalfi\"achen, Mathematische Werke \textbf{3}, Verlagsbuch-handlung, Hillesheim, 219--248








http://iopscience.iop.org/1751-8121/49/30/305201

  




\end{thebibliography}
\end{document}